\documentclass[10pt,times,mathptm,psfig,twocolumn,conference]{IEEEtran}
\usepackage{graphicx}  % Written by David Carlisle and Sebastian Rahtz

\usepackage{subfigure} % Written by Steven Douglas Cochran
\begin{document}

\title{Spatially resolved THz response as a characterization concept for nanowire FETs}

\author{\authorblockN{Klaus Michael Indlekofer}
\authorblockA{Institute for Bio and Nanosystems, CNI\\
Research Center J\"ulich\\
D-52425 J\"ulich, Germany\\
Email: m.indlekofer@fz-juelich.de} \and
\authorblockN{Radoslav N\'emeth}
\authorblockA{Institute for Bio and Nanosystems, CNI\\
Research Center J\"ulich\\
D-52425 J\"ulich, Germany} \and
\authorblockN{Joachim Knoch}
\authorblockA{IBM Research GmbH\\
Zurich Research Laboratory\\
8803 Rueschlikon, Switzerland}}

% avoiding spaces at the end of the author lines is not a problem with
% conference papers because we don't use \thanks or \IEEEmembership

%Homer Simpson\authorrefmark{2},
%Montgomery Scott\authorrefmark{3} and
%Eldon Tyrell\authorrefmark{4}}
%Telephone: (800) 555--1212, Fax: (888) 555--1212}
%\authorblockA{\authorrefmark{4}Tyrell Inc., 123 Replicant Street, Los Angeles, California 90210--4321}}

%\specialpapernotice{(Invited Paper)}

\maketitle

\begin{abstract}
In this paper, we propose a THz probe technique to obtain spatially
resolved information about the electronic spectra inside
nanowire-based FETs. This spectroscopic approach employs a segmented
multi-gate design for the local detection of quantum transitions
between few-electron states within the FET channel. We simulate the
intra-band THz response of such devices by means of a many-body
quantum approach, taking quantization and Coulomb interaction
effects into account. The obtained simulation results demonstrate
the capabilities of the proposed technique which go beyond the
limitations of standard characterization methods.
\end{abstract}


% for peerreview papers, inserts a page break and creates the second title.
% Will be ignored for other modes.
\IEEEpeerreviewmaketitle

\section{Introduction}

The electronic properties of ultimately scaled nanotransistors are
dominated by only a handful of electrons or holes. In such
nanodevices, one has to face two non-classical physical mechanisms.
Firstly, due to the spatial confinement of charge carriers on a
nanoscale, quantization energies of electronic states become
relevant, leading to a non-classical transport behavior. Secondly,
the details of the Coulomb interaction between the individual
carriers become important and can no longer be described in terms of
a classical mean-field picture. The consequences can be clearly
observed experimentally for example in Coulomb blockade phenomena.

One-dimensional nanowire-based field effect transistors (FET) have
recently attracted great interest due to their advantageous
electrostatics and transport properties \cite{thela,guo,indle05}.
From a different perspective, they also represent prototypes for the
study of technological as well as physical challenges in future
transistor designs. Instead of considering quantization and Coulomb
phenomena as unwanted deviations from the classical device behavior,
here one actually utilizes such non-classical effects for the
improvement of FET devices or for new functionalities in a future
quantum information technology.

In order to understand and finally utilize the physics of
quantization and Coulomb effects within such nanowire FETs, one
needs a characterization technique that is able to probe quantum
states. Typical quantum energy scales of few-electron states in
realistic nanowire FETs are on the order of a few meV corresponding
to the 100GHz-1THz frequency range. We therefore propose the concept
of a spatially resolved THz probe to directly measure quantum
transitions between few-electron (or -hole) states within a nanowire
FET. This spectroscopic approach employs a multi-gate FET layout,
thus being able to provide spatially resolved information about the
electronic spectra and electronic charge distributions inside the
FET channel. Such a concept goes beyond the capabilities of standard
linear and nonlinear characterization methods \cite{charFET}. In
this paper, we discuss the theoretical aspects of the proposed
multi-gate THz probe. We simulate the intra-band THz response of
nanowire FET devices by means of a many-body quantum approach,
taking quantization and Coulomb interaction effects beyond
mean-field into account.

In the following sections, we first describe the main physical ideas
behind our many-body THz simulation approach, and secondly, we
demonstrate the strengths of the concept of a THz probe by
discussing a realistic example of a multi-gate nanowire FET. As a
main effect, we obtain the formation of a Wigner molecule in the
long channel limit.

\begin{figure}
\centering
\includegraphics[width=3.2in]{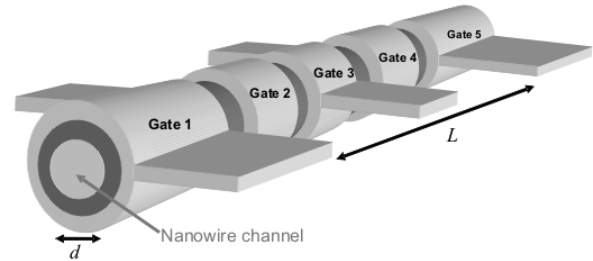}
\caption{Schematic view of an idealized nanowire MOSFET in
multi-gate configuration (``caterpillar FET''). This example shows a
system with 5 gates. Here, gates 1 and 5 are negatively biased and
thus provide barriers. Gates 2,3,4 cover a channel region of length
$L$ and act as probes. In the simulations, the inter-gate gaps are
assumed to be negligible. The outer source and drain contacts are
not shown here.} \label{fig1}
\end{figure}

\section{Theoretical approach}

Fig.~\ref{fig1} shows a schematic sketch of the considered nanowire
FET with multiple gate segments. We assume that only one lateral
subband needs to be taken into account, corresponding to channel
diameters in the sub-30 nm range for GaAs. In order to capture the
numerous details of a real device structure, such as the actual
geometrical configuration of the gate electrodes, a realistic
quantum simulation of such a nanodevice requires the consideration
of a sufficiently large number $Z\approx 50-500$ of single-particle
states \cite{lake,pul}. In contrast to the commonly employed
Anderson- or Hubbard-like many-body models \cite{anderson,hubbard},
in general we have to account for an inhomogeneous, anisotropic, and
screened Coulomb interaction. Since the resulting many-body problem
scales exponentially in $Z$, such a realistic simulation thus grows
beyond any computational limit. In this context, we have recently
introduced a multi-configurational approach (MCSCG,
\cite{indle05,indle06,indle07}) which employs a reduced adaptive
basis for the simulation of Coulomb blockade phenomena in nanowire
FETs. For the simulation of THz response, however, the number of
considered basis states has yet to be increased. As a solution, we
have therefore extended the MCSCG by a ``bucket-brigade'' algorithm
(BBCI) \cite{indlecondmat06,indle07x} which allows us to select a
few thousand relevant many-body configurations from the
exponentially growing basis set. A general few-electron state in
turn can be written as a linear combination of such relevant
configurations. This approach enables us to describe correlated
many-body states beyond mean-field theory (e.g. Hartree-Fock or
LDA-DFT \cite{DFT}) which become essential within the field of
quantum information technology. The following simulation results are
based on the BBCI algorithm in combination with first-order Kubo
theory \cite{kubo} for the description of the THz response spectra
within the chosen relavant many-body basis.

\section{Simulation results}

For the following idealized example, the outer ends of the channel
are covered by gate electrodes (for example gates 1 and 5 in
Fig.~\ref{fig1}). Applying a sufficiently large negative voltage to
the outermost gates, barriers arise and we thus can assume an almost
isolated channel system. The input THz excitation is applied to gate
1 (see Fig.~\ref{fig1}), whereas the intermediate gate fingers
(gates 2,3,4 in Fig.~\ref{fig1}) serve as spatially resolved probes
with zero DC bias. In our example, we consider two different
GaAs-based nanowire MOSFETs with SiO$_2$ shell and $40+2$ gate
segments. In both cases we keep the channel diameter $d=20$nm and
the oxide thickness $d_{ox}=10$nm constant, whereas the channel
lenghts are chosen as $L=150$nm and $L=600$nm with corresponding
total oxide capacitances of $C_{ox}=47$aF and $C_{ox}=188$aF,
respectively. Due to the coaxial gate electrodes, the Coulomb
interaction within the channel region is strongly screened. In the
considered example, we obtain a screening length \cite{auth,indle05}
of $\lambda=10.5$nm for $\epsilon_{ox}=3.9$ and
$\epsilon_{ch}=12.4$. In the following discussion, we assume a
thermodynamical equilibrium state in the low temperature limit in
order to obtain a well-resolved occupation of quantum states. The
total electron number is kept fixed at $N=4$, trapped within the
inner channel region $L$. As for a possible experimental preparation
of such an electronic state, one could employ the outer source and
drain contacts in combination with suitable barrier-gate voltages to
make use of single-electron tunneling for electron counting
\cite{been}.

%\begin{figure*}
%\centerline{ \subfigure[(a)]
%{\includegraphics[width=3.0in]{dens150NoEE.eps}
%\label{fig:dens150NoEE}} \hfil \subfigure[(b)]
%{\includegraphics[width=3.0in]{kubo150NoEE.eps}
%\label{fig:kubo150NoEE}}} \caption{Simulated GaAs nanowire FET with
%$L = 150$ nm and $d = 20$ nm for the case {\it without}
%electron-electron interaction. The number of electrons is $N = 4$.
%(a) Electron charge density. (b) Top-view of the nanowire FET (40+2
%gates) and spatially resolved THz transition spectrum. Within the
%grayscale plot, white corresponds to a strong response signal.}
%\label{fig:150NoEE}
%\end{figure*}

\begin{figure}
\centering
\includegraphics[width=3.3in]{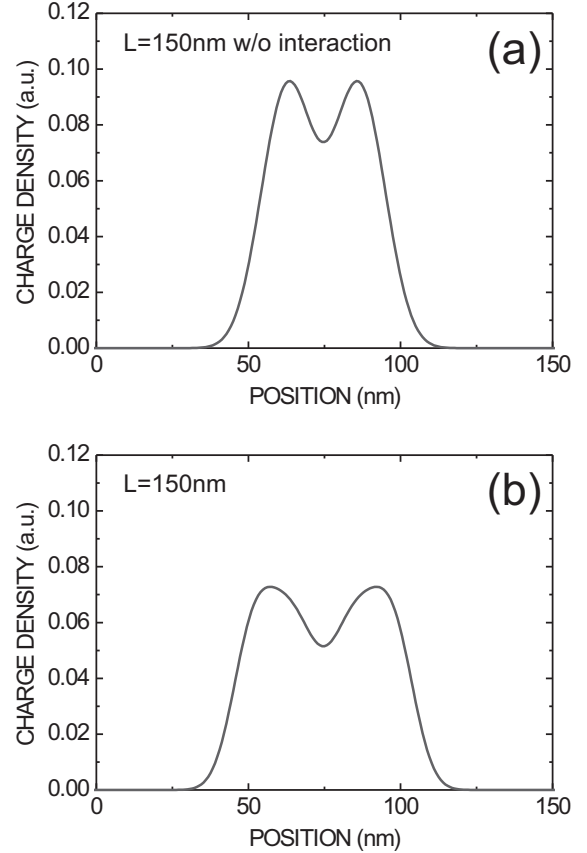}
\caption{Simulated electron charge density for a GaAs nanowire FET
with $L=150$nm and $d=20$nm. The number of electrons is $N=4$. (a)
shows the non-interacting case, whereas (b) includes the
electron-electron interaction. Comparing the two cases, no
significant qualitative change can be observed.} \label{fig:dens150}
\end{figure}

\begin{figure}
\centering
\includegraphics[width=3.3in]{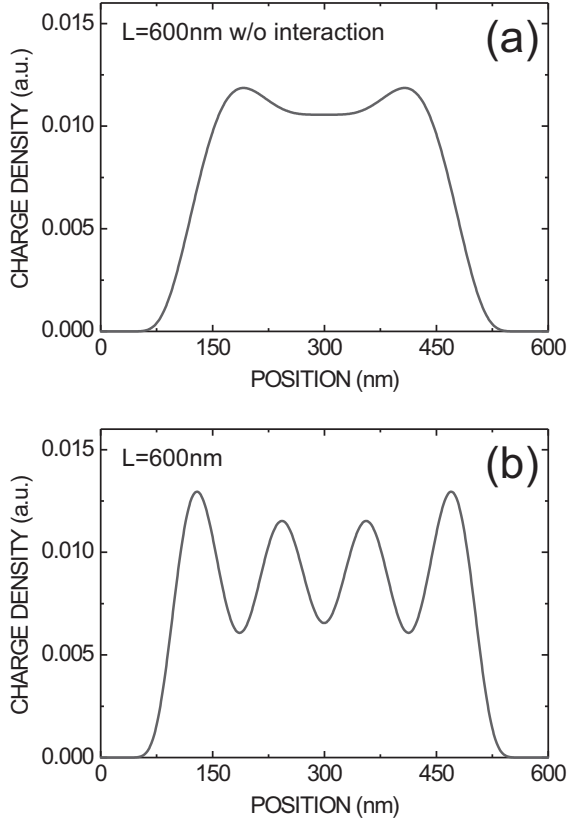}
\caption{Simulated electron charge density for a GaAs nanowire FET
with $L=600$nm and $d=20$nm. The number of electrons is $N=4$. (a)
shows the non-interacting case, whereas (b) includes the
electron-electron interaction. The charge-density wave in (b)
indicates the formation of a Wigner-molecule within the long-channel
device. Here, the electron density differs qualitatively from the
non-interacting case (a).} \label{fig:dens600}
\end{figure}

Fig.~\ref{fig:dens150} shows the simulated charge density along the
channel axis for the nanowire FET with $L=150$nm. For comparison,
the non-interacting case is shown in Fig.~\ref{fig:dens150}(a),
whereas the realistic case which includes the Coulomb repulsion is
plotted in Fig.~\ref{fig:dens150}(b). For this case, the
single-particle quantization energy dominates as compared to the
Coulomb repulsion, resulting in a charge distribution which
resembles the shape of a non-interacting system. Here, the Coulomb
repulsion solely broadens the spatial electron distribution. In
contrast, for the $L=600$nm case (Fig.~\ref{fig:dens600}(a) and (b))
one can clearly identify the formation of a charge density wave,
indicating the onset of the Wigner molecule regime \cite{haeusler}
with separated electrons owing to the dominating Coulomb repulsion.
The actual transition to the Wigner regime for a nanowire FET in
general results from a competition between kinetic energy and
Coulomb energy, which of course depends on the chosen materials and
geometries. In the long-channel case ($L=600$nm), the mean spatial
separation of the electrons $\Delta x\simeq L/N$ becomes much larger
than the screening length $\lambda$ of the Coulomb interaction
within the channel. For the discussed example, in turn, it is
energetically favorable for the electrons to form a charge density
wave in order to reduce the Coulomb energy, outweighing the
increased kinetic energy due to the spatial confinement $\Delta x$.
Comparing the non-interacting case Fig.~\ref{fig:dens600}(a) with
Fig.~\ref{fig:dens600}(b), one can clearly observe a qualitative
change in the electron density profile.

In order to probe such a peculiar electronic configuration within a
nanowire FET channel, we now consider the spatially resolved THz
response spectrum of such a device. In the following simulation
results, we have employed a resolution of 40 intermediate probe
gates. One has to note that each individual gate finger detects a
spatially averaged signal within an effective interval
$l_G^{eff}\simeq l_G+2\lambda$, where $l_G$ denotes the geometrical
length of the gate segment and $\lambda$ the gate screening length.
Parasitic stray-capacitances which will likely occur in an
experimental realization of such a device are not considered in this
paper. For the experimental case it might be advantageous to employ
alternating screening and signal gates, combined with an optimized
THz layout.

%\begin{figure}
%\centering
%\includegraphics[width=3.2in]{kubo150NoEE.eps}
%\caption{Top-view of the nanowire FET (40+2 gates) and the simulated
%THz transition spectrum for a GaAs nanowire with $L = 150$ nm and
%$d = 20$ nm for the case {\it without} electron-electron
%interaction. Within the grayscale plot, black corresponds to a
%strong response signal. The number of electrons is $N = 4$.}
%\label{fig:kubo150NoEE}
%\end{figure}

\begin{figure}
\centering
\includegraphics[width=3.3in]{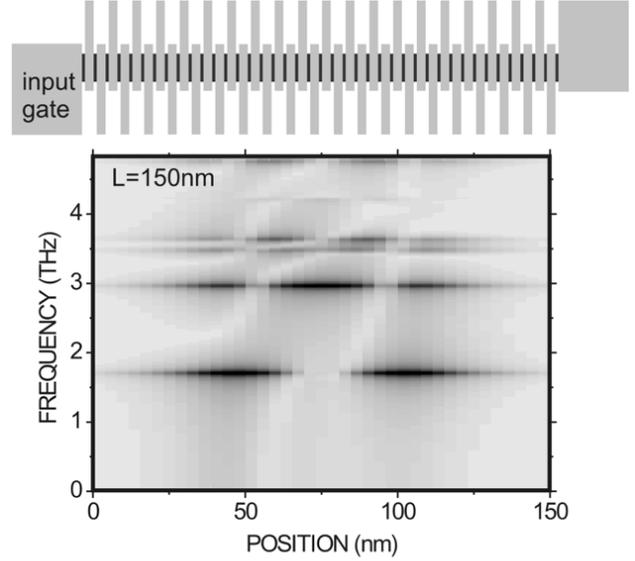}
\caption{Top-view of the nanowire FET (40+2 gates) and the simulated
THz transition spectrum for a GaAs nanowire with $L = 150$ nm and $d
= 20$ nm. The number of electrons is $N = 4$. Within the grayscale
plot, black corresponds to a strong response signal. For this
device, the main effect of the electron-electron interaction
consists mainly in a renormalization of the energy spectrum.}
\label{fig:kubo150}
\end{figure}

%\begin{figure}
%\centering
%\includegraphics[width=3.3in]{kubo600NoEE.eps}
%\caption{Top-view of the nanowire FET (40+2 gates) and the simulated
%THz transition spectrum for a GaAs nanowire with $L = 600$ nm and
%$d = 20$ nm for the case {\it without} electron-electron
%interaction. Within the grayscale plot, black corresponds to a
%strong response signal. The number of electrons is $N = 4$.}
%\label{fig:kubo600NoEE}
%\end{figure}

\begin{figure}
\centering
\includegraphics[width=3.4in]{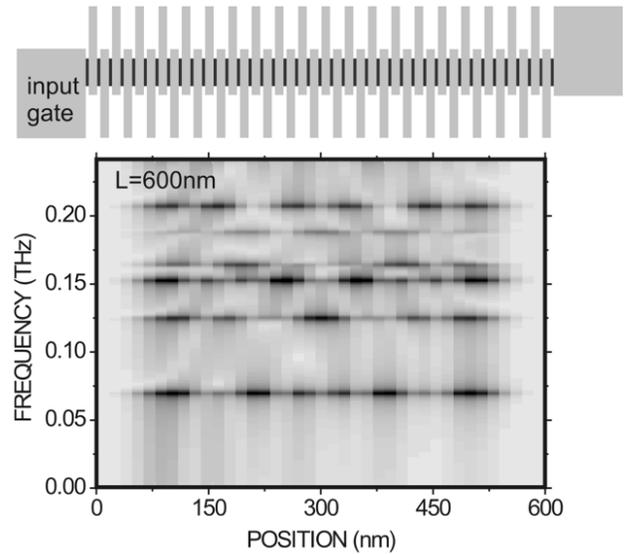}
\caption{Top-view of the nanowire FET (40+2 gates) and the simulated
THz transition spectrum for a GaAs nanowire with $L = 600$ nm and $d
= 20$ nm. The number of electrons is $N = 4$. Within the grayscale
plot, black corresponds to a strong response signal. As compared to
the short-channel case in Fig.~\ref{fig:kubo150}, the formation of a
Wigner-molecule leads to additional spatial peaks.}
\label{fig:kubo600}
\end{figure}

Fig.~\ref{fig:kubo150} and Fig.~\ref{fig:kubo600} show the simulated
THz spectra for the two cases $L=150$nm and $L=600$nm, respectively,
which exhibit fundamental resonances at 1.69THz and 70GHz. The
higher value for the 150nm case stems from a narrower electron
distribution (see Fig.~\ref{fig:dens150}(b)). Most noticeably,
comparing the qualitative form of the two spectra, signatures of the
Wigner-like regime for the 600nm case (Fig.~\ref{fig:kubo600}) can
be identified via the appearance of additional spatial peaks which
are related to the oscillatory nature of the charge density wave.
One has to note that the maxima in these transition spectra are
located at those gate positions where the charge {\it oscillates}
the most, and thus need not coincide with the charge density maxima.
In fact, for a non-interacting case, the spatial THz pattern is
related to the gate-averaged {\it product} of the two wavefunctions
of the involved final and destination states \cite{indle07x}. Even
with the inclusion of the electron-electron interaction, the
considered $L=150$nm device comes close to such a situation. In the
case of a correlated many-body state, however, such an
interpretation is not applicable in general and a many-body approach
such as the BBCI becomes mandatory. In any case, the obtained THz
spectra can be considered as ``fingerprints'' of the concrete
electronic configuration of the channel, providing not only
information about electronic transition energies but also about the
spatial configuration of the underlying few-electron states.

\section{Conclusion}

In summary, we have considered a THz probe for a spatially resolved
analysis of electronic spectra in nanowire-based transistors
employing a multi-segment gate design. We have simulated the THz
response of few-electron quantum states in nanowire FETs by use of a
recently developed numerical many-body technique. The discussed
example of a GaAs-based device demonstrates that signatures of
Wigner-like charge density waves can be identified by use of this
method, which lies beyond the scope of standard FET characterization
methods. As such, the proposed multi-gate THz probe technique might
prove useful in a future experimental realization as a means to
characterize nanoscale devices which are dominated by quantization
and Coulomb effects.

\end{document}